\documentclass[sigconf,
]{acmart}

\renewcommand\footnotetextcopyrightpermission[1]{} 
\usepackage{booktabs} 
\usepackage{balance}
\usepackage{subcaption}
\usepackage{graphicx}
\usepackage{caption}
\setcopyright{rightsretained}

\acmConference[TheWebConf'18]{The Web Conference}{April 2018}{Lyon, France} 
\acmYear{2018}
\copyrightyear{2018}

\begin{document}
\title{Measuring bot and human behavioral dynamics}

\author{Iacopo Pozzana}
\affiliation{%
  \institution{Birkbeck, University of London}
  \city{London} 
  \state{UK}
}
\additionalaffiliation{%
  \institution{USC Information Sciences Institute}
  \streetaddress{4676 Admiralty Way, Suite 1001}
  \city{Marina Del Rey} 
  \state{CA} 
}
\email{iacopo@dcs.bbk.ac.uk}

\author{Emilio Ferrara}
\affiliation{%
  \institution{USC Information Sciences Institute}
  \streetaddress{4676 Admiralty Way, Suite 1001}
  \city{Marina Del Rey} 
  \state{CA} 
}
\email{emiliofe@usc.edu}


\begin{abstract}
Bots, social media accounts controlled by software rather than by humans, have recently been under the spotlight for their association with various forms of online manipulation. To date, much work has focused on social bot detection, but little attention has been devoted to the characterization and measurement of the behavior and activity of bots, as opposed to humans'. Over the course of the years, bots have become more sophisticated, and capable to reflect some short-term behavior, emulating that of human users. The goal of this paper is to study the behavioral dynamics that bots exhibit over the course of one activity session, and highlight if and how these differ from human activity signatures.
By using a large Twitter dataset associated with recent political events, we first separate bots and humans, then isolate their activity sessions. We compile a list of quantities to be measured, like the propensity of users to engage in social interactions or to produce content.
Our analysis highlights the presence of short-term behavioral trends in humans, which can be associated with a cognitive origin, that are absent in bots, intuitively due to their automated activity.
These findings are finally codified to create and evaluate a machine learning algorithm to detect activity sessions produced by bots and humans, to allow for more nuanced bot detection strategies.
\end{abstract}

%
%

\ccsdesc[500]{Networks~Social media networks}
\ccsdesc[500]{Information systems~Web and social media search}

\keywords{social bots, bot behavior, human dynamics, social media \& society}

\maketitle

\section{Introduction}
Over the course of the last few years, social media have been conveying
an ever increasing portion of human communications.
At the same time, research progress and increasing availability of cheap
hardware has brought the emergence of sophisticated forms of artificial intelligence.
The concurrence of these two factors is at the root of the emergence of social
bots, both as a field of investigation for the scientific community and as a
topic of interest for the generalist media and the society at large \cite{ferrara2016rise, adams2017ai}.

Social bots are all those social media accounts that are controlled by
artificial, as opposed to human, intelligence.
Their purposes can be many: news aggregators, for example, collect and relay
pieces of news from different sources; chatbots can be used as automated
customer assistants.
However, as a by now large number of studies has shown,
the vast majority of bots are employed as part of large-scale efforts
to manipulate public opinion or sentiment on social media,
such as for viral marketing or electoral campaigns, often with 
quantifiable effects ~\cite{forelle2015political, bessi2016social, woolley2016automating}.

In light of this fact, scholars' efforts to investigate social bots
can roughly be grouped in two categories.
On one side, many studies have focused on the theme of bot detection, i.e.,\
on how to recognize bot accounts as such \cite{chu2012detecting, clark2016sifting, davis2016botornot}.
A second line of research deals
instead with the impact of the viral deployment of automated accounts on
large-scale social phenomena such as information spread and sentiment manipulation \cite{howard2016bots, shao2017spread}.

The characterization of bots' behaviour is thus a topic that can yield actionable insights, especially
when considered in comparison with humans'.
Our present work examines an aspect of such comparison that, to the best of our
knowledge, has not yet been explored in detail in the existing literature:
the short-term behavioural dynamics, i.e.,\ the temporal evolution of behavioural
patterns over the course of an activity session of bots as opposed to that of humans.
Prior studies have examined  the performance
of human users when engaging in continuous online interactions, finding measurable
changes, for example, in the amount of reactions to other users' post,
or in the quality (in terms of grammatical correctness and readability) of the produced content   \cite{singer16reddit,kooti2016behavioral}.

We hypothesize that such behavioural changes, if at all present, should be
starkly different in the case of bot accounts, when compared with human counterparts.
To investigate the matter, we analyze a dataset of posts from the Twitter
platform, focusing our attention on the discussion preceding the 2017
French presidential election. A previous study considered the role played by bot accounts in that
context, finding evidence of the presence of a large number of such actors,
working to create (or destroy) consensus around specific candidates by means of
(mis)information diffusion---a phenomenon also observed in many other analogous events.

\subsection{Contributions of this work}

Over the course of single activity sessions, we measure
different quantities capturing user behaviour, e.g.,\
propensity to engage in social interactions, or amount of produced content,
and finally contrast results between bots and humans.

The present study advances our understanding of bots and human user behavior in the following ways:

\begin{itemize}
    \item We reveal the presence of short-term behavioural trends among humans that are instead absent in the case of bots. Such trends may be explained by a deterioration of human user's performance (in terms of quality and quantity of produced content), and by an increasing engagement in social interactions over the course of an online session; in both cases, we would not expect bots to be affected, and indeed we record no significant evidence of any short-term temporal evolution for this category of users.

    \item In the spirit of the research line on bot detection, we codify our findings in a set of highly predictive features capable of separating human activity sessions from bots' ones; then, we design and evaluate the performance of a machine learning framework that leverages this features to detect bot activity sessions. 
    This can prove extremely desirable when trying to detect so-called \textit{cyborgs}, users that are in part controlled by humans and in part bots. Our session classification system yields an accuracy of 94\% AUC (\textit{Area Under the ROC curve}). The addition of the features identified by our analysis yields an average improvement over the baseline of up to 14\% AUC.

\end{itemize}

\section{Data \& Methods} \label{data}
Our dataset consists of a collection of more than 16M tweets,
posted by more than 2M different users.
The tweets were posted between April 25 and May 7, 2017, the two-weeks period
leading to the second round of the  French presidential election.
A list of 23 keywords and hashtags was manually compiled and used to collect the data
through the Twitter Search API.\footnote{https://dev.twitter.com/rest/public/search}

To classify the users between bots and humans, we employed the \emph{Botometer} API\footnote{https://botometer.iuni.iu.edu/}
(previously known as BotOrNot \cite{davis2016botornot}),
which provides a free-to-use, feature-based classification system.
When queried about a Twitter user name or user id, Botometer retrieves from Twitter
information about more than a thousand features associated with that account,
grouped into six categories: content, friend, network, sentiment, temporal, and user meta data;
the API returns a \emph{bot score} for each category, plus an overall score.
A bot score is a number representing the likelihood for the account to be controlled
by a bot, and it ranges from 0 (definitely human) to 1 (definitely bot).

Here, we used Botometer to calculate the bot score of more than 380k users in our dataset,
namely all users who posted at least 5 tweets during the observation time,
minus those whose account was since deleted (27k), or which privacy setting
prevented Botometer to access the necessary information (15k accounts).
The 380k users are responsible for more than 12M out of the overall 16M tweets. 

It is important to note that Botometer does not use any session-related features, nor does incorporate any notion of activity sessions \cite{varol2017online}: this is important to guarantee that the behavioral differences that later emerge are not just an artifact of the classifier using session-based features to separate bots from humans (that would be circular reasoning).

\begin{figure}
 \centering
  {\includegraphics[width=0.98\columnwidth]{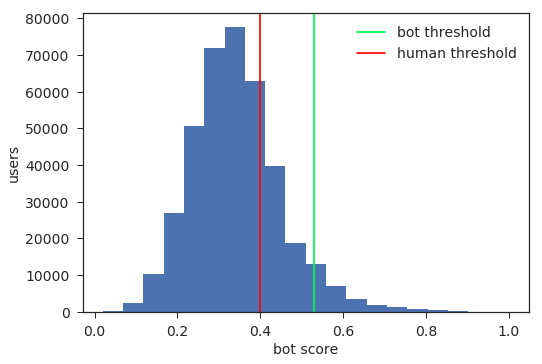}}
  \caption[bot score]
  {
  Frequency distribution of the bot scores obtained with Botometer.
  We classified as bots the 5\% top scoring accounts,
  corresponding to a minimum value of 0.53 (green line).
  We set instead 0.4 as a threshold (maximum) value for the labeling of a human.
  Both choices are informed by previous analyses and intended to reduce the risk of misclassifications \cite{varol2017online}.
  }
  \label{botscore}
\end{figure}

The distribution of the bot scores is reported in Figure \ref{botscore}.
To limit the risk of wrongly classifying a human account,
we chose to only label as bots those users with a bot score ranking in the top 5\%
of the distribution, corresponding to a threshold value of 0.53. This is a conservative strategy informed by the fact that a false positive, i.e., labelling a human user as a bot, is generally associated to a higher cost than a false negative, i.e., mislabelling a bot as human, especially when decisions (such as account suspensions) are informed by this classification. Furthermore, recent analyses by the Botometer's authors demonstrated that, when studying human and bot interactions, results do not significantly vary in the threshold range between 0.4 and 0.6~\cite{varol2017online}.
According to the same conservative strategy, we set the threshold for humans to 0.4, leaving unlabeled all the accounts with a score value
between the two thresholds. Summarizing, we have 19k users labeled as bots and 290k users labeled as humans, while the reminding 78k
are left unlabeled.


To organize the dataset in sessions, the tweets were first grouped by user and sorted
according to the time of posting.
A session is a group of consecutive tweets separated by an amount of time larger than a certain threshold
of $T$ minutes, or in other words, every time an user posts a tweet after a period of inactivity of at least
$T$ minutes, we say that s/he (or it) has started a new sessions.

\begin{figure}
 \centering
  {\includegraphics[width=0.98\columnwidth]{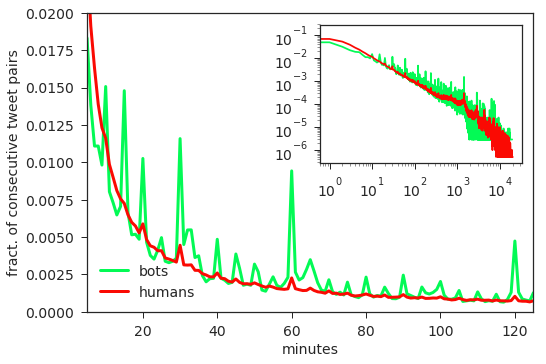}}
  \caption[inter-tweet time]
  {
  Distribution of the time difference between two consecutive tweets.
  The overall distribution (insect) shows the characteristic long tail
  for both humans and bots. The detail of the 10 minutes to 2 hours time range (main figure)
  reveals the presence of peaks corresponding to multiples of 5 minutes;
  the peaks are present both in the humans' and bots' distribution,
  but they are much more marked in the latter.
  }
  \label{intert}
\end{figure}

To determine the value of $T$ we first considered the distribution of the inter-time between two consecutive
tweets from the same user, reported in Figure \ref{intert}.
The overall distribution (\textit{cf.} inset of Figure \ref{intert}) displays the characteristic long tail,
both for humans and bots. In human behaviour, this is a common feature, known as burstiness \cite{goh2008burstiness};
observing burstiness among bots does not come as a surprise either,
as the newer, most sophisticated bots are indeed known to sample their inter-event times
from long-tailed distributions, precisely for the purpose of avoiding detection~\cite{ferrara2016rise,ferrara2017disinformation}.
However, a closer inspection, centred on the typical time range of a session
duration (10 minutes to 2 hours, main figure), highlights the presence of peaks
corresponding to regular values (10 minutes, a quarter of an hour, half an hour
and so forth) and, although the peaks are present in both distributions,
they are significantly more pronounced in the case of bots.

\begin{figure}
 \centering
  {\includegraphics[width=0.98\columnwidth]{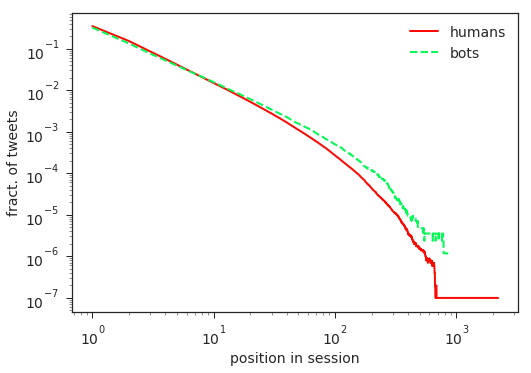}}
  \caption[inter-tweet time]
  {
  Fraction of tweets appearing in a given position in the course of a session. Both distributions (bots and humans)
  are long-tailed, with the first showing higher values in the tail, thus indicating that bot are likelier than humans
  to post more tweets in the course of a same session (i.e. without a break of 60 minutes or more).
  }
  \label{sessions}
\end{figure}

Also informed by previous studies \cite{halfaker2015session, singer16reddit}, we set a threshold value of 60 minutes
for our analysis. Additional results, not reported here, obtained with smaller
threshold values (10, 15 and 30 minutes), show the same qualitative trends.
The resulting number of sessions is more than 250k for the bots and 2.6M for the humans.
The frequency distribution of the tweets according to their position in the session, plotted normalized in Figure \ref{sessions}
to highlight the comparison between humans and bots, is long tailed, and shows how bots are more likely to
engage in longer sessions (with a number of tweets in the order of the tens or larger).

\section{Experimental Analysis} \label{exp}
\begin{figure*}
        \begin{subfigure}{0.49\textwidth}
                \includegraphics[width=\linewidth]{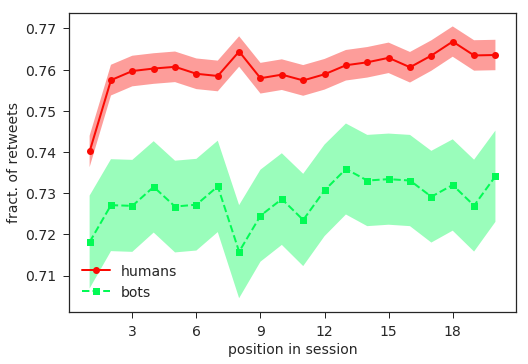}
                \caption{Fraction of retweets}
                \label{ret-pos}
        \end{subfigure}%
        \begin{subfigure}{0.49\textwidth}
                \includegraphics[width=\linewidth]{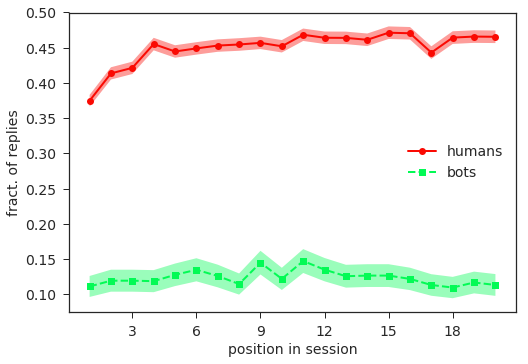}
                \caption{Fraction of replies}
                \label{rep-pos}
        \end{subfigure}%
        \\
        \begin{subfigure}{0.49\textwidth}
                \includegraphics[width=\linewidth]{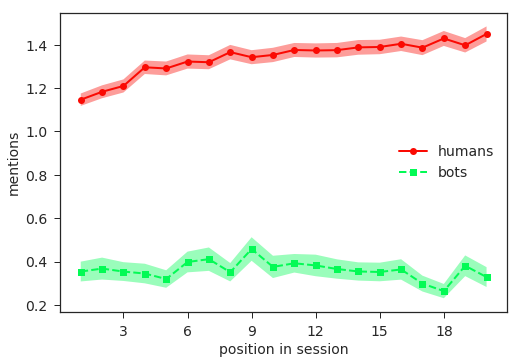}
                \caption{Mentions per tweet}
                \label{men-pos}
        \end{subfigure}%
        \begin{subfigure}{0.49\textwidth}
                \includegraphics[width=\linewidth]{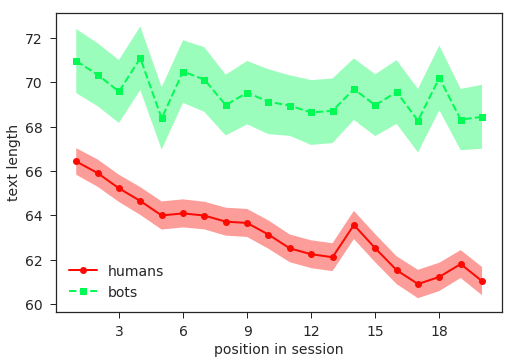}
                \caption{Text length}
                \label{text-pos}
        \end{subfigure}%
        \caption{Trend of four different behavioural measures over the course of an online session.
        All the sessions considered here contain a similar number of posts (20 to 25) to limit biases due to
        the different behavioural patterns adopted by users in sessions of different length \cite{singer16reddit}.
        The number of sessions considered is thus 1500 for the bots and 13k for the humans.
        With the obvious exception of frame \ref{ret-pos}, retweets have been excluded from our analysis,
        as we are only interested in the original contend produced by an user. The shaded area corresponds to one SEM,
        calculated separately for each point.
        For all measures, not only the datapoints are well separated between the two categories of users,
        but humans also show temporal trends which are not observed in bots. In particular:
        considering the fraction of retweets (\ref{ret-pos}), values are higher for human users with
        respect to bots over all the course of a session;
        humans also show an increase in their value, faster in the beginning (first 2-3 tweets),
        then slower, but still present during all the first 20 tweets.
        The situation is similar in the case of replies (\ref{rep-pos}).
        Human users also use more mentions (\ref{men-pos}), with a roughly steady increase over the course of a session.
        Coming to the average length of the tweets (\ref{text-pos}), a decrease is evident for humans, who also
        post shorter tweets with respect to their automated counterparts.
        In all the four measures considered, no clear trend emerges for the case of bots.
        }
        \label{experiments}
\end{figure*}

Once classified the users as bots and humans, and organized the tweets in sessions,
we proceed to study the temporal dynamics of the two categories of users.
Our results are summarized in Figure \ref{experiments}. We focus on four quantities:
the fraction of retweets and of replies among all tweets posted at a certain position in a session;
the number of mentions appearing in a tweet; and the length of the tweet itself.

As detailed below, the first three of these four features provide an indicator of the
quantity and quality of the social interactions of an user over the course of a session.
The fourth one (text length) is instead a measure of the amount of content produced by an user.
As correlations between the length of a session and the dynamics of performance indicators
have been observed on social networks \cite{kooti2016behavioral, singer16reddit},
we carry out our analysis on sessions of similar lengths only (20 to 25 posts),
resulting in a total of 1500 bot sessions and 13k human sessions.

A retweet is a repost of a tweet previously posted by another user.
We thus expect to see an increase in the number of human retweets during the course
of a session, as users get exposed to other users' posts.
The fraction of retweets over the total number of tweets, grouped by their position in the session,
is shown in Figure \ref{ret-pos}: in general, the fraction of retweets is higher for humans at all positions;
human users increase their number of retweets over all the course of their sessions, starting with
a rapid increase in the first 2-3 posts and then slowing down.
Such trend is not evident among bots, that seem instead to oscillate around a constant value.

As a second type of interaction, we consider the reply. The reply, as the name suggests,
is a tweet posted in response to some other tweet. The same considerations as for the retweets apply here:
we expect to see the fraction of replies increase over the course of a human sessions.
Our results, reported in Figure \ref{rep-pos}, confirm our expectation: as for the retweets,
the fraction of replies increases and decelerates, for humans, over all
the first 20 tweets. Bots, on the other hand, don't exhibit an analogous increase.

On Twitter, users can mention other users in their post; 
another possible measure of social interactions is thus the average number of mentions per post.
As for the previous cases, we expect the number of mentions to increase, on average, as human
users proceed in their session. The results (Figure \ref{men-pos}) do indeed show an increase
in the average number of mentions by humans over the course of the first 20 tweets.
Again, bots don't seem to change their behaviour in the course of the session.

The features analyzed so far are all indicators of the amount of social interactions in which users engage.
We now consider the average length (in characters) of a tweet, which is a measure of the amount of content produced
and is thus an interesting indicator of the short-term behavioural dynamics.
Before, counting the number of characters, the tweet is stripped off all urls, mentions and hashtags,
so to only account for the part of the text effectively composed by the user.
A previous study has failed to show any significant variation in this quantity over the course of a short-term
session on Twitter \cite{kooti2016behavioral}; however, analyses of other platforms have shown that
the average post length decreases on similar time scales \cite{singer16reddit}.
Here, the human data show a clear decreasing trend, whereas no trend emerges for what concerns the bots
(see Figure \ref{text-pos}).

Notice that for the last three quantities (replies, mentions and text length) we have excluded
all retweets from our analysis, as their content is not produced by their poster:
whereas the fact itself of posing a retweet can be considered a behavioural indicator,
the content of the retweet could hardly provide any valuable insight in this respect.

In general, our experiments reveal the presence of a temporal evolution in the human behaviour over the
course of a session on an online social network, whereas, confirming our expectations,
no evidence is found of a similar evolution for bot accounts. In the next session,
we proceed to further investigate the significance of these temporal trends by incorporating
them in a classifier for bot detection.

\section{Predictions} \label{pred}
As the experiments described in the previous section show,
human behaviour over the course of an online session evolves in a measurably
different manner with respect to bot users.
To further investigate this difference, we implement a classifier that,
leveraging the quantities considered above, categorizes tweets as either produced by
a bot or a human. Using five different off-the-shelf machine learning algorithms,
we train our classifier using 10-fold cross-validation on a dataset of labeled
tweets; the dataset, different from the one analyzed above, consists in three
groups of tweets produced bot account active in as many viral spamming campaigns at
different times, plus a group of human tweets. All accounts are labeled by human annotators,
and a thorough description of the dataset is provided in \cite{cresci2017paradigm}.
In particular, the groups of accounts considered here are the ones named 'social spambots \#1-3'
and 'genuine accounts' \cite{cresci2017paradigm}, for a total of circa
3.4M tweets posted by 5k bot accounts and 8.4M tweets posted by 3.5k human accounts.

We proceed to organize the dataset in session separate by 60 minutes intervals,
as described in Section \ref{data} above. As a result, each tweet is tagged with
three \emph{session features}: session id, position of the tweet in the session and length of the session.
Six \emph{behavioural features} are also considered: whether the tweet is a retweet, or a reply,
the numbers of mentions, hashtags and urls contained in the tweet, and the text length.
We use the nine features to train four classifiers, using five different techniques:
Decision Trees, Extra Trees, Random Forest and Adaptive Boosting.

\begin{figure*}
        \begin{subfigure}{0.49\textwidth}
                \includegraphics[width=\linewidth]{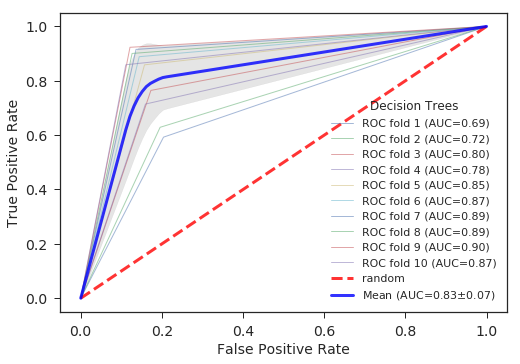}
                \caption{Decision Trees}
                \label{dt}
        \end{subfigure}%
        \begin{subfigure}{0.49\textwidth}
                \includegraphics[width=\linewidth]{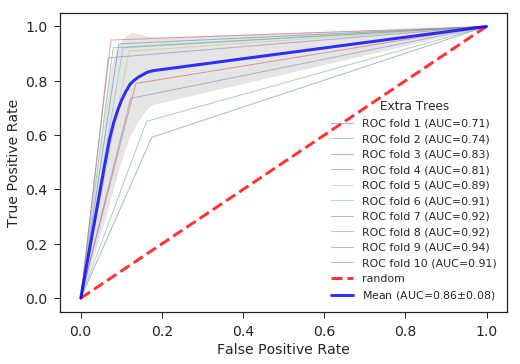}
                \caption{Extra Trees}
                \label{et}
        \end{subfigure}%
        \\
        \begin{subfigure}{0.49\textwidth}
                \includegraphics[width=\linewidth]{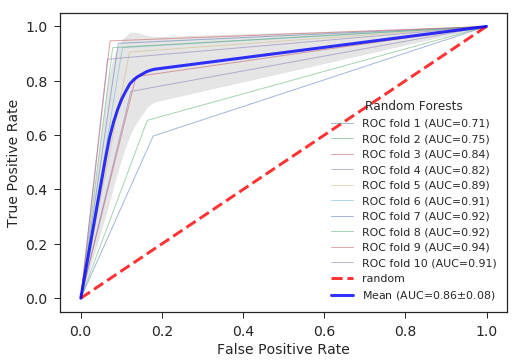}
                \caption{Random Forests}
                \label{rf}
        \end{subfigure}%
        \begin{subfigure}{0.49\textwidth}
                \includegraphics[width=\linewidth]{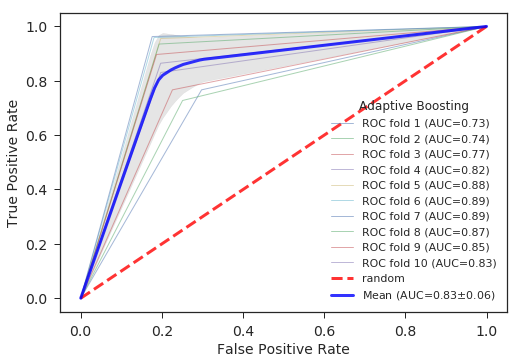}
                \caption{Adaptive Boosting}
                \label{ab}
        \end{subfigure}%
        \\
        \begin{subfigure}{0.49\textwidth}
                \includegraphics[width=\linewidth]{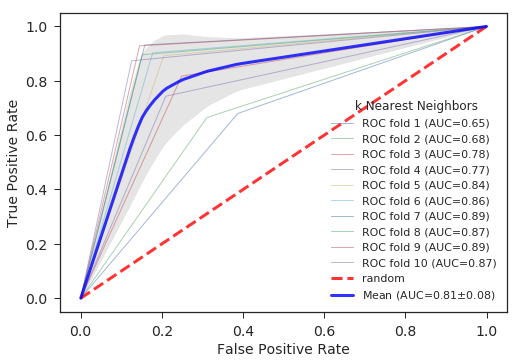}
                \caption{k Nearest Neighbors}
                \label{knn}
        \end{subfigure}%
        \caption{Cross-validation AUC plots for five specific models: \ref{dt} Decision Trees (DT),  \ref{et} Extra Trees (ET),
         \ref{rf} Random Forests (RF),  \ref{ab} Adaptive Boosting (AB) and \ref{knn} k Nearest Neighbors (kNN).
         For all the four models, the plot shows ROC curves and AUC scores (in legend) for each of the 10 cross-validation folds (thinner lines),
         and the mean ROC curve (blue line) with its AUC (in legend) and confidence area ($\pm$ 1 std, in gray).
         ET and RF yield the best cross-validated average performance (Mean AUC=86\%), followed by DT and AB (83\%) and kNN (81\%).}
         \label{crossval}
\end{figure*}

The training and testing of the model is done via 10-fold cross-validation on the entire dataset.
The details are reported in Figure \ref{crossval}; on average over the 10 folds,
the AUC scores between 86\% and 81\% for all models.

\begin{figure*}
\begin{subfigure}{0.45\textwidth}
    \includegraphics[width=\linewidth]{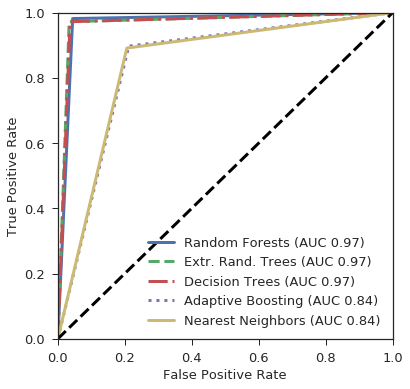}
    \caption{Full model}
    \label{full}
\end{subfigure}
\begin{subfigure}{0.45\textwidth}
    \includegraphics[width=\linewidth]{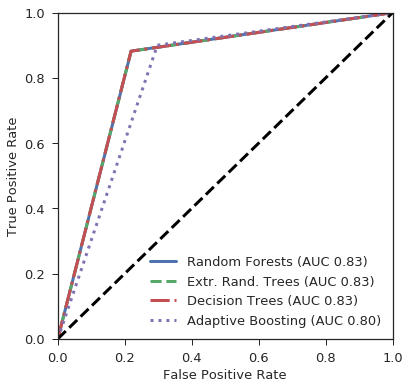}
    \caption{Baseline}
    \label{base}
    \end{subfigure}
    \caption{Area Under the ROC curve plot. We show five  models, yielding the highest AUC among the trained ones. Each of the models exploits three session features in conjunction with six behavioural features. These models all significantly outperform the baseline models that do not include the three session features derived from our experimental analysis.}
    \label{roc}
\end{figure*}

As a further verification, for each one of the five models, we choose the best classifier among the ten obtained in the
cross-validation and test it on the full dataset. The performance, as expressed by the AUC measure,
improves for the DT, the ET and the RF, raising from 90\%, 94\% and 94\% respectively to 97\%, and decreasing for the
Ab and the kNN, dropping from 89\% to 84\%; the ROC curves are plotted in Figure \ref{full}.
However, aside from the details of the effectiveness of each classifier,
the results just described go to show that short-term behavioural patterns can effectively be used to inform
bot detection.

To quantify precisely the impact of the introduction of the information concerning the session dynamics,
we train four more classifiers, equivalent to the ones described above in all respects,
except for the set of features used for the training: here only the behavioural features
(retweet, reply, hashtags, mentions, urls, text length) are included while the three features
characterizing the session dynamics (session id, position in session, session length)
are left out. The four models (again DT, ET, RF and AB; we omitted kNN, the slowest to train
among all these models, due to time constraints) are again trained and tested via 10-fold cross-validation.
We don't report here the details of all the folds, but again we further tested the best classifier for each method
on the whole dataset, and the corresponding ROC curves are plotted in Figure \ref{base}.
The new four models serve as a baseline to compare the full models to, in order
to get a quantitative indication of the impact of the introduction of the three session features
on the performance of the classifiers.
The difference is particularly pronounced for the first three models (DT, ET, RF),
for which the AUC yields a 83\% for the baseline versions, 14 points lower than their
counterparts trained with all the nine features. The AB model also performs worse
without the session features (AUC 80\%, compared to the 84\% obtained with the
full features).

All the testing of our classifiers was done, until this point,
on the annotated dataset from \cite{cresci2017paradigm}.
As an example of the effectiveness of leveraging session-level behavioural dynamics for bot detection,
we would now be interested in carrying out some sort of testing on the
dataset of the French election tweets introduced in Section \ref{data}.
As such dataset lacks annotations, a proper test
can not be performed, but we can still exploit the Botometer scores to
get some information about the performance of our classifiers,
and again draw a comparison with the baseline case where session features are omitted.
To this purpose, we let the bot threshold (Botometer score value above which
an account is consider a bot) on all the range of values between 0 and 1,
and for each case compare the results given by the classifiers, trained
on the 'spambots' dataset as described above, with these 'annotations'.
Again, let us remark that our purpose here is to evaluate the effectiveness
of the introduction of the feature describing the session dynamics,
and not to exactly evaluate the sensitivity of the classifiers.

\begin{figure}
 \centering
  {\includegraphics[width=0.98\columnwidth]{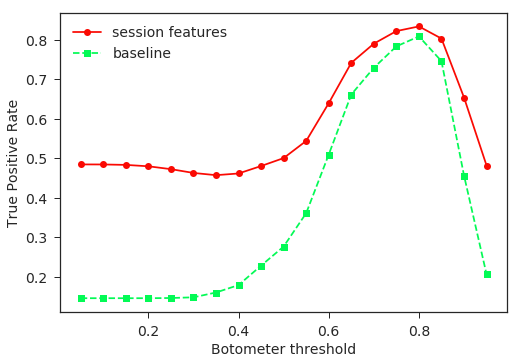}}
  \caption[inter-tweet time]
  {
   Comparison of the performance on the French election data between the two AB classifiers trained on the 'spambots' data.
   One classifier is trained with the full features (red circles), while in the training of the other
   the session features are omitted (green squares). The set of 'positives' is taken as all the tweets
   posted by accounts scoring more than the corresponding Botometer score (horizontal axis), and could
   indeed contain some human accounts, especially for lower values. Nonetheless, the higher TPR goes to show
   that the introduction of the session features significantly improves the performance of the classifier.
  }
  \label{comp}
\end{figure}

The test is performed using the two AB classifiers, the one trained
with the full features and the baseline one, and the results are shown in Figure \ref{comp}.
The left part of the graph is not actually very informative, as when the bot
threshold is set below 0.4, the 'positive' account will actually include many humans.
It is roughly in correspondence of the 0.4 value that the true positive rate (TPR)
of the classifier starts increasing , and although the baseline classifier's
TPR increases as well, the former outperforms the latter at all points.

To summarize, the results exposed in this section show that  features describing
the short-term behavioural dynamics of the users can effectively be employed to implement a
bot detection system or to improve existing ones, thus further confirming that a difference
exists in such dynamics between humans and bots.

\section{Discussion}
The results detailed in the previous two Sections provide evidence of the existence
of significant differences in the temporal evolution of behaviour over the course of an online
session between human and bot users. 

In particular, in Section \ref{exp} we analyze four different indicators of the
users' behaviour and find, among humans, trends that are not present among bots:
first of all, an increase in the amount of social interaction, measured as the fraction
of retweets, the fraction of replies, and the number of mentions contained in a tweet;
secondly, a decrease in the amount of content produced, measured as the average tweet length.
Such trends are present up to the 20th post in human sessions,
whereas the same indicators remain roughly constant for bots.
This may be partly due to the fact that, as a sessions progresses, users grow more tired
and become less likely to perform more complex actions such as composing an entirely
original post \cite{kooti2016behavioral}.
At the same time, we hypothesize that another possible (and possibly concurring) explanation
may be given by the fact that, as time goes by, users are exposed to more and more
posts, thus increasing their probability to react, for example by retweeting or by mentioning
the author of a previous post.
In both cases, bots would not be affected by such considerations, and no behavioural
change should be expected from them.

It is worth noting again that Botometer does not implement any notion of activity sessions nor does it use any session-based features for bot classification \cite{varol2017online}. This ensures that the behavioral differences highlighted in this work are genuine and not simply an artifact due to discriminating on features used for classification purposes (that would be circular reasoning).

In Section \ref{pred}, we use the results obtained in Section \ref{exp}
to inform a classification system for bot detection. Our purpose there is to
highlight how the introduction of features describing the session dynamics
(session id, position of the tweet in the session and length of the session)
can substantially improve the performance of the detector.
To this purpose, we use a range of different machine learning techniques
(Decision Trees, Extra Trees, Random Forests, Adaptive Boosting, k Nearest Neighbors)
to train, through 10-fold cross-validation, two different set of classifier:
one including the features describing the session dynamics (the full model), and
one without those features (the baseline). The comparison between the two sets of models,
carried out both on the annotated dataset user for the cross-validation and on the
dataset of tweets concerning the French elections, show that the full model significantly outperforms
the baseline.

\section{Related work}

Although bots in some rare occasions have been used for social good, e.g., to deliver positive interventions and interactions~\cite{savage2016botivist,monsted2017evidence}, their use is mostly associated with malicious operations. Bots, for example, have been involved with social media manipulation of political conversation~\cite{metaxas2012social,forelle2015political,howard2016bots,woolley2016automating,bessi2016social}, the spread of disinformation and fake news~\cite{ferrara2017disinformation,shao2017spread}, conspiracy~\cite{subrahmanian2016darpa} and extremist propaganda~\cite{ferrara2016predicting,FERRARA20171}, as well as stock market manipulation~\cite{ferrara2015manipulation}.

The increasing evidence brought our research community to propose a wealth of techniques to address the challenges posed by the pervasive presence of bots in platforms like Facebook and Twitter. Social bot detection is one such example. A recent review~\cite{ferrara2016rise} suggested to classify bot detection approaches under three classes: \emph{(a)} methods based on social network structure and dynamics; \emph{(b)} systems based on crowd-sourcing and human annotations; \emph{(c)} learning algorithms based on informative features that separate bots from humans. Our work differentiates from this literature as it is not directly aimed at bot detection, yet our findings can be used to inform detection  based on bot and human features and behaviors.

The study of bots' characteristic is another recent research thread that attracted much attention. Researchers discovered that bots exhibit a variety of diverse behaviors, capabilities, and intents~\cite{mitter2014categorization,varol2017online}. A recent technical memo illustrated novel directions in bot design that leverage Artificial Intelligence (AI): AI bots can generate media and textual content of quality potentially similar to human-generated content but at much larger scale, completely automatically~\cite{adams2017ai}. In this work, we highlighted similarities and dissimilarities between bots' and humans' behavioral characteristics, illustrating the current state of bots' capabilities.

Provided that bots oftentimes operate in concert (botnets), this attracted the attention of the cybersecurity research community. Examples of such botnets have been revealed on Twitter~\cite{abokhodair2015dissecting,zhou2017starwars}. Botnet detection is still in its early stage, however much work  assumed unrestricted access to social media platform infrastructure. Different social media providers, for example, applied bot detection techniques in the back-end of other platforms, like Facebook~\cite{stein2011facebook,beutel2013copycatch} and Renren (a chinise Twitter-like social platform)~\cite{wang2013you,yang2014uncovering}. Although these approaches can be valuable and show promising results~\cite{cao2012aiding,stein2011facebook,alvisi2013sok}, for example to detect large-scale bot infiltration, they can be implemented exclusively by social media service providers with full access to data and system infrastructure. 

Researchers in academic groups who don't have unrestricted access to social media data and systems, proposed many alternative techniques that can work well with smaller samples of user activity, and fewer labelled examples of bots and humans. The research presented here is one such example. Other examples  include the classification system proposed by Chu \textit{et al.}~\cite{chu2010tweeting,chu2012detecting}, the crowd-sourcing detection framework by Wang \textit{et al.}~\cite{wang2012social}, the NLP-based detection methods  by  Clark \textit{et al.}~\cite{clark2016sifting}, and the BotOrNot classifier~\cite{davis2016botornot}. 

Some historical user activity data is still needed for these methods to function properly, either by indirect data collection ~\cite{chu2010tweeting,lee2011seven,chu2012detecting,wang2012social,clark2016sifting}, or, like in the case of BotOrNot~\cite{davis2016botornot}, by interrogating the Twitter API (which imposes strict rate limits, making it impossible to do large-scale bot detection). Given these limits, we believe that it is very valuable to have a deep understanding of human and bot behavioral performance dynamics: our findings can inform data collection and annotation strategies, can help improve classification accuracy by injecting expert knowledge and produce better, more informative and predictive features, and ultimately allow for a better understanding of interaction mechanisms online.

\section{Conclusions} 
In the present work we have investigated the behavioral dynamics of social network
users over the course of an online session, with particular attention to the differences
emerging between human and bot accounts under this perspective.
User session dynamics have been investigated in the literature before but,
to the best of our knowledge, never applied to the problem of bot detection.

Our analysis revealed the presence of behavioral trends at the session level
among humans that are not observed in bot accounts.
We hypothesized two possible mechanisms motivating such trends: on one side,
humans' performance deteriorates as they engage in prolonged online sessions; this decline has been attributed to a cognitive origin in related work.
On the other hand, over the course of their online activity, humans are constantly exposed
to posts and messages by other users, so their probability to engage in social
interaction increases. Devising methods to further test each of these two hypotheses
could possibly constitute an avenue for future research.
Furthermore, the presence of such behavioral differences between the two categories of users
can be exploited to improve bot detection techniques. To investigate this possibility,
we trained two categories of classifiers, in one case including features describing the
session dynamics, while omitting them in the other. The comparison shows that session
features bring an increase of up to 14\% AUC, thus substantially improving the performance
of the bot detectors. This suggests that features inspired by cognitive dynamics can be useful indicators of human activity signatures, which may be harder to replicate by bots. Importantly, the adopted classifier does not leverage any session-related features, thus ensuring that the results we observe are genuine and not the artifact of circular reasoning.
It may be an interesting object of future work to better characterize
the interplay between the features studied here and other features leveraged by various bot detection techniques, such
as the ones mentioned in the "Related Work" Section.

Overall, our study contributes both to the ongoing investigation around the detection
and characterization of social bots, and to the understanding of online human behaviour,
specifically in its short-term dynamical evolution over the course of activity sessions.

\begin{acks}

\small{The authors thank Kristina Lerman (USC) for insightful discussions, Stefano Cresci and collaborators (IIT-CNR) for sharing the bot annotations, and the Botometer team (IU) for maintaining their public bot detection tool.

\smallskip

\noindent The authors gratefully acknowledge support by the Air Force Office of Scientific Research (AFOSR, award number FA9550-17-1-0327), and by the Defense Advanced Research Projects Agency (DARPA, contract number W911NF-17-C-0094, and grant number D16AP00115). The U.S. Government is authorized to reproduce and distribute reprints for Governmental purposes notwithstanding any copyright annotation thereon. The views and conclusions contained herein are those of the authors and should not be interpreted as necessarily representing the official policies or endorsements, either expressed or implied, of AFOSR, DARPA, or the U.S. Government.}
\end{acks}

\balance

\bibliographystyle{ACM-Reference-Format}
\bibliography{bib} 

\end{document}